\documentclass[12pt]{article}

\usepackage[dvips]{graphicx}
\usepackage{epsfig}
\usepackage{amsmath}
\usepackage{amssymb}
\usepackage{amsfonts}

\usepackage{amsmath,amssymb,pictex,cite,eepic,epsfig,psfrag,url}
\advance \topmargin by -\headheight
\advance \topmargin by -\headsep
\evensidemargin \oddsidemargin
\makeatletter
\@addtoreset{equation}{section}

\makeatother

\makeatletter
\def\Appendix{\appendix
  \def\@seccntformat##1{Appendix~\csname the##1\endcsname.~~}}
\makeatother

\begin{document}
\hfill \hbox{ITEP-LAT/2010-09}
\vspace{1.5cm}

\bigskip

\begin{center}
{\Large \textbf{Torus Amplitudes in Minimal Liouville Gravity and Matrix Models}}

\vspace{1.0cm}

{\large V.~Belavin}

\vspace{0.2cm} Theory Department of Lebedev Physical Institute and \\
Institute of Theoretical and Experimental Physics\\ Moscow, Russia  

\end{center}

\vspace{1.0cm}

\textbf{Abstract}
We evaluate one--point correlation numbers on the torus in the Liouville theory coupled to the 
conformal matter $M(2,2 p+1)$. We find agreement with the recent results obtained in the matrix model
approach.

\section{Introduction}
The two-dimensional Liouville gravity~\cite{Polyakov} remains one of the very few consistent quantum field theories
involving dynamical metric field. General formulation of the Liouville gravity, whose action is induced
by critical matter, allows one to consider a special type of ``solvable'' examples,
in which the matter sector is represented by some minimal model~\cite{BPZ} of $2D$ conformal field
theory. We use the term ``minimal Liouville gravity'' for such models.
Since long ago it is believed that the scaling limit of matrix models (see e.g.~\cite{GinsM}
and references therein)
gives an alternative description of the minimal Liouville gravity. Nevertheless, at present, a proof of this
statement is still missing. In this situation it seems to be desirable to improve the understanding
of the relations between these two approaches.
In~\cite{Moore,BZ} a 
way to identify the results of the matrix models with those of the minimal Liouville gravity was 
found for the conformal matter 
represented by the non-unitary series $(2,2p+1)$ of the CFT minimal models. In~\cite{BZ} a resonance transformation, 
which relates the coupling parameters of the Liouville gravity with the 
couplings of the matrix models, was constructed. In terms of the transformed parameters the matrix models
correlation numbers 
should  coincide with the naturally defined correlation numbers in the framework of the minimal Liouville gravity.
Recently~\cite{Eynard,BT,BBT}, the problem of matrix model analysis in higher genera was revisited. In particular, 
in~\cite{BT} the torus contribution
to the generating function of one-matrix models was found and the resonance transformation was applied to find  one- and 
two-point correlation numbers on the torus in the Liouville frame. 
The aim of this paper is to test the matrix models results available from~\cite{BT} against
direct calculations in the minimal Liouville gravity.

\section{Minimal Gravity $\mathcal{MG}_{2/2p+1}$}
The main problem of the minimal Gravity is to construct and to evaluate the gravitational correlation
functions. In the Polyakov approach~\cite{Polyakov} the functional 
integral over metrics is reduced to the moduli integral over Riemann surfaces.
The integrand involves the correlation functions of the ghosts $b,c$ and the vertex 
operators $U_k=\Phi_{k} V_{a_k}$ constructed by an appropriate Liouville dressing of the matter fields.
Due to the factorized form of the vertex operator the integrand of the moduli integral splits into the product 
of the Matter, the Liouville and the ghosts correlation functions. All three theories are conformally invariant.
The central charge of the Liouville theory
\begin{equation}
c^{\text{L}}=1+6Q^2
\end{equation}
where $Q=b+b^{-1}$ and $b$ is the Liouville coupling, is related to the central charge of the conformal matter 
by means of the so called central charge balance condition $c^M+c^L=26$, which is equivalent to 
the requirement of the total Weyl (BRST) invariance. In the minimal Liouville 
gravity $\mathcal{MG}_{p/q}$ the matter sector is described by the CFT minimal model
$M(p/q)$ with the central charge 
\begin{equation}
c^{\text{M}}=1-\frac{6(p-q)^2}{pq}
\end{equation}
which possesses a set of primary fields $\Phi_{m,n}$ with $m\in(1,\cdots,p-1)$ and $n\in(1,\cdots,q-1)$
of conformal dimensions 
\begin{equation}
\Delta^{\text{M}}_{m,n}=\frac{(np-mq)^2-(p-q)^2}{4pq}
\end{equation} 
The central charge balance condition determines the value of the Liouville coupling to be $b=\sqrt{p/q}$.
The conformal dimension 
\begin{equation}
\Delta^{\text{L}}(a)=a(Q-a)
\end{equation}
 of the exponential 
Liouville field $V(a)$ in the construction of 
the vertex operator $U_{m,n}=\Phi_{m,n}  V(a)$ is also fixed by the Weyl invariance, which 
requires that the total conformal dimension 
of the vertex operator $\Delta^{\text{M}}_{m,n}+\Delta^{\text{L}}(a)=1$. This yields $a=a_{m,-n}$,  
where
\begin{equation}
a_{m,n}=\frac{(1-m) b^{-1}+(1-n) b}{2}
\end{equation} 

In the torus $N$-point amplitude the conservation of the ghost current requires
one vertex insertion to be fixed~\cite{Polchinski}. In order to be BRST invariant, 
this insertion should be decorated by ghost 
fields as follows
\begin{eqnarray}
\langle O_{k_1} O_{k_2} \cdots O_{k_N} \rangle_{\text{torus}}= \int_{F} d\tau d\bar{\tau}
\langle b(0)\bar b(0) c(0) \bar c(0) U_{k_1}(0) \prod_{n=2}^N 
\int d z_n d \bar z_n U_{k_n}(z_n,\bar z_n)\rangle_{\tau} 
\label{npoint}
\end{eqnarray}
Here $F$ is the fundamental region of the modular group: $\tau_2>0, |\tau|>1,-1/2\leq\tau_1<1/2$.
The expectation value at the right hand side involves the matter, the Liouville and the ghost sectors
considered on the torus with the modular parameter $\tau=\tau_1+i \tau_2$.
In what follows we are interested in the minimal Liouville gravity $\mathcal{MG}_{2/2p+1}$.
Then the Liouville coupling constant is 
\begin{equation}
b=\sqrt{\frac{2}{2p+1}}
\label{bp}
\end{equation}
an the one-point amplitude reads
\begin{eqnarray}
\langle O_{k} \rangle_{\text{torus}} = \int_{F} d\tau d\bar{\tau}
\langle b \bar b c \bar c \rangle_{\tau}\langle \Phi_{k} \rangle_{\tau}\langle V(a_{1,-k-1})\rangle_{\tau} 
\label{1pointB}
\end{eqnarray}
where we used the brief notation $\Phi_{k}=\Phi_{1,k+1}$ and $k=0,\cdots,p-1$. 
The 4-point correlation function in the ghost sector is given by~\cite{Polchinski}
\begin{eqnarray}
\langle b \bar b c \bar c \rangle_{\tau}=|\eta(q)|^4
\end{eqnarray}
with $\eta(q)=q^{1/24} \prod_{k=1}^{\infty} (1-q^k)$ being the Dedekind eta function and 
$q=e^{2 i \pi \tau}$. In terms of the CFT on the complex plain the one-point correlation functions on the
torus with the modular parameter $\tau$ takes the form  
\begin{equation}
\langle \Phi_k \rangle_{\tau}= \text{Tr} (q \bar q)^{L_0-c^{\text{M}}/24} \Phi_k =
\sum_{\{\Delta\}} C_{\Delta_k, \Delta}^{\Delta} 
(q \bar q)^{\Delta-c^{\text{M}}/24} |F^{\text{M}}(\Delta_k,\Delta,q)|^2 
\label{1pointM}
\end{equation}
\begin{equation}
\langle V_a \rangle_{\tau}= \text{Tr} (q \bar q)^{L_0-c^{\text{L}}/24} V_a =
\int \frac{d P}{4\pi} C_{a,Q/2+i P}^{Q/2+i P}
(q \bar q)^{\Delta(P)-c^{\text{L}}/24} |F^{\text{L}}(\Delta(a),\Delta(Q/2+i P),q)|^2 
\label{1pointL}
\end{equation}
Here $C_{\Delta_k, \Delta}^{\Delta}$ and $C_{a,Q/2+i P}^{Q/2+i P}$ are the 
structure constants of the operator algebras in the Matter and the Liouville sectors correspondingly,
while $F(\Delta_{\text{ext}},\Delta_{\text{int}},q)$ is the one-point conformal block function defined 
as the contribution of the highest weight representation of the Virasoro algebra with the conformal dimension 
$\Delta_ {\text{int}}$. 
In~\cite{FatLet,Poghos} recursive relations for the Liouville conformal block function were found, which 
make it possible
to calculate its expansion into a power series of $q$. In~\cite{Leshek} the crossing symmetry for this 
representation was checked numerically.  

Consider first the most simple example $\langle O_0 \rangle$. 
The structure constant  $C_{0,\Delta}^{\Delta}=\delta_{0,\Delta}$ so that in the matter sector we
just have the partition function of the corresponding minimal model. It is expressed
in terms of the characters of the irreducible representations of the Virasoro algebra.
It is interesting that in this case the dressing Liouville correlation function $\langle V(a_{1,-1}) \rangle_{\tau} $ 
can be evaluated explicitly. 
The external conformal dimension in~\eqref{1pointL} turns to be 
$\Delta_{\rm ext}=\Delta^{\text{L}}(a_{1,-1}=b)=1$. 
Using  the recursive 
algorithm proposed in~\cite{FatLet} we verified up to $20$th order in $q$ that the
conformal block $F(1,\Delta_{\text{int}},q)$ does not depend on the internal conformal dimension
$\Delta_{\rm int}$ and is equal to
\begin{equation}
F(1,\Delta_{\text{int}},q)=\frac{q^{1/24}}{\eta(q)}
\end{equation}  
The general expression for the diagonal Liouville structure constant is 
\begin{eqnarray}
C_{a,Q/2+i P}^{Q/2+i P}=(\pi \gamma(b^2) b^{2-2b^2})^{-a/b} 
\frac{\Upsilon(b)\Upsilon(2a)\Upsilon(2 i P)\Upsilon(-2 i P)}
{\Upsilon^2(a)\Upsilon(a+2 i P)\Upsilon(a-2 i P)}
\label{StrCnstLiouv}
\end{eqnarray}
Using the definition of the Upsilon function (see e.g.~\cite{ZZ3point}), for $a=b$ we find
\begin{equation}
C_{b,Q/2+i P}^{Q/2+i P}=\frac{4 P^2}{\pi b} 
\end{equation}
One can perform the $P$ integration in~\eqref{1pointL} analytically. This yields
\begin{eqnarray}
\langle O_0 \rangle_{\text{torus}} =\frac{1}{4\pi^2 b} 
\int_F d \tau\, d\bar \tau \,\tau_2^{-3/2} |\eta(q)|^2 \sum_{s=1,\cdots,p} |\chi_{1,s}(q)|^2
\label{1pointE}
\end{eqnarray}
where the characters of the irreducible representations  explicitly read 
(see e.g~\cite{YellowBook})
\begin{equation}
\chi_{1,s}(q)=\frac{q^{\frac{(2s-(2p+1))^2}{8(2p+1)}}}{\eta(q)}
\sum_{k\in Z}\bigg(q^{2(2p+1)k^2+(2s-(2p+1))k}-q^{2(2p+1)k^2+(2s-(2p+1))k+s}\bigg)
\end{equation}
The result can be written as
\begin{eqnarray}\label{1pointF}
\langle O_0 \rangle_{\text{torus}}= 
\frac{1}{4\pi^2 b}
 \sum_{i=1,\cdots,p} 
\sum_{m,n\in Z} \bigg( I(\alpha_{n,i},\alpha_{m,i},\delta_i)\nonumber\\-2I(\alpha_{n,i},\beta_{m,i},\delta_i)
+I(\beta_{n,i},\beta_{m,i},\delta_i)\bigg)
\end{eqnarray}
where
\begin{equation}
\alpha_{k,i}=2(2p+1) k^2 +k(2i-2p-1)
\end{equation}
\begin{equation}
\beta_{k,i}=2(2p+1) k^2 +k(2i+2p+1)+i\\
\end{equation}
\begin{equation}
\delta_i=\frac{(2i-2p-1)^2}{8(2p+1)}
\end{equation}
and
\begin{eqnarray}
I(\alpha,\beta,\delta)=\int_{-1/2}^{1/2} d x e^{2\pi i(\alpha-\beta)x} \int_{\sqrt{1-x^2}}^\infty
d y y^{-3/2} e^{-2 \pi(2 \delta+\alpha+\beta)y}
\end{eqnarray}

It turns out that the correlation number $\langle O_0 \rangle$ can be evaluated analytically. The 
calculation is based on the following ideas~\cite{klebanov}. The torus partition function
of the minimal model $M(2,2p+1)$ is related to the torus partition function of a free scalar field as
\begin{equation}
Z_{M(2/(2p+1))}=\sum_{s=1,\cdots,p} |\chi_{1,s}(q)|^2=
\frac12 \bigg[Z_B\bigg(g=\sqrt{2(2p+1)}\bigg)-Z_B\bigg(g=\sqrt{2/(2p+1)}\bigg)\bigg]
\end{equation}
where 
\begin{eqnarray}
Z_B(g)=\frac{1}{|\eta(q)|^2}\sum_{s,t} q^{(s g^{-1}+t g)/4}\bar q^{(s g^{-1}-t g)/4}
\end{eqnarray}
By using the Poisson resummation formula one can derive that
\begin{eqnarray}
Z_B(g)=g \frac{1}{\sqrt{\tau_2}|\eta(q)|^2}\sum_{n,m}e^{-\frac{\pi g^2|n-m\tau|^2}{\tau_2}}
\end{eqnarray}
This form of the matter partition function allows one to calculate~\eqref{1pointE} explicitly
\begin{eqnarray}
\langle O_0 \rangle_{\text{torus}} =\frac {1}{8\pi^2 b} 
\bigg[J\bigg(\sqrt{2(2p+1)}\bigg)-J\bigg(\sqrt{2/(2p+1)}\bigg)\bigg]
\end{eqnarray}
where
\begin{eqnarray}
J(g)=g\int_F \frac{d^2 \tau}{\tau_2^2}\sum_{n,m}e^{-\frac{\pi g^2|n-m\tau|^2}{\tau_2}}
=g \bigg\{\int_F\frac{d^2 \tau}{\tau_2^2}+2\sum_{k=1}^{\infty}\int_{-1/2}^{1/2} d\tau_1 \int_0^{\infty}\frac{d \tau_2}{\tau_2^2} e^{-\frac{\pi g^2 k^2}{\tau_2}}\bigg\}
\label{J}
\end{eqnarray}
where the $m$-summation in the second term is replaced by the sum over inequivalent images of the fundamental 
region, which together cover the strip $-\frac 12 \leq \tau_1 \leq \frac 12$. Performing the integrations
in~\eqref{J} we find
\begin{eqnarray}
J(g)=\frac\pi 3 \bigg(g+\frac1 g\bigg)
\end{eqnarray}
The result for the one-point amplitude takes the simple form 
\begin{eqnarray}
\langle O_0 \rangle_{\text{torus}} =\frac {p}{24\pi}
\label{1pointexact}
\end{eqnarray}
We checked numerically for 
different values of $p$ that even if we only retain the first term under the sum over $(m,n)$ 
in~\eqref{1pointF}, the results~\eqref{1pointF} and~\eqref{1pointexact} 
coincide with the several digits precision. To compare it with the results of the matrix models 
we will need to consider the ratio of two one-point amplitudes. This allows us to avoid the problem of a different 
normalizations of the partition functions.
In this verification it turns out to be sufficient to evaluate the correlation numbers 
$\langle O_k \rangle_{\text{torus}}$ only
retaining the zero-order terms in the conformal block expansions.
In the given approximation the matter correlation function reads
\begin{eqnarray}
\langle \Phi_{k}\rangle_{\tau} =\sum_{m=1,\cdots, p} C_{\Delta_{1, k+1},\Delta_{1,m}}^{\Delta_{1,m}}
|q|^{2\Delta_{1,m}-\frac{c^M}{12}}
\end{eqnarray}
and the dressing Liouville function looks like
\begin{eqnarray}
\langle V(a_{1,-k-1}) \rangle_{\tau} = \int \frac{d P}{4\pi} C_{(1+k/2)b,Q/2+i P}^{Q/2+i P} 
|q|^{2\Delta_P-\frac{c^L}{12}}
\end{eqnarray}
We can present the zero-order result for the one-point correlation number~\eqref{1pointB} in the following form
\begin{eqnarray}
\langle O_k \rangle_{\rm torus} =\sum_{m=1,\cdots, p} C_{\Delta_{1, k+1},\Delta_{1,m}}^{\Delta_{1,m}} I_{k,m}
\label{Ok}
\end{eqnarray}
where
\begin{eqnarray}
I_{k,m}&=&\int_F d^2 \tau \int \frac{d P}{4\pi}\,  C_{(1+k/2)b,Q/2+i P}^{Q/2+i P} |q|^{2(\alpha_{m}+P^2)}
\nonumber\\
&=&\frac{1}{\pi}\int_{0}^{1/2} d x 
\int_{0}^{\infty} d P\, \frac{C_{(1+k/2)b,Q/2+i P}^{Q/2+i P}}{\alpha_{m}+P^2}
e^{-4\pi (\alpha_{m}+P^2) \sqrt{1-x^2}}
\end{eqnarray}
and
\begin{eqnarray}
\alpha_m=\frac{Q^2}{4}-\frac{c^{\text{L}}+c^{\text{M}}}{24}+\Delta_{1,m}
=\frac{(1+2(p-m))^2}{8(1+2p)}
\end{eqnarray}
For the comparison with the matrix models results we
 will need explicit expressions for
the structure constants for $k=1,2$. The minimal model structure constant for $k=1$  reads
\begin{eqnarray}
 C_{\Delta_{1, 2},\Delta_{1,m}}^{\Delta_{1,m+1}}=
\bigg(\frac{\gamma(2-2\rho)\gamma(1-m\rho)}{\gamma(1-\rho)\gamma(2-(1+m)\rho)}\bigg)^{\frac 12}
\label{C12}
\end{eqnarray}
where $\rho=2/(2p+1)$. Notice that the degenerate field $\Phi_{1,2}$ have no diagonal 
channels in the operator product expansion. Thus, naively the correlation number~\eqref{1pointB}  
for $k=1$ (as well as for any odd $k$) should vanish, which, in particular, 
contradicts the results 
of the matrix models. The solution of this contradiction is rather   
simple. Taking into account the symmetry of the Kac table for $M(2,2p+1)$ one can see that the operators $\Phi_{1,p}$
and $\Phi_{1,p+1}$ have the same conformal dimension and thus represent the same physical field.
Hence, in the case $k=1$ the only nonvanishing term is that with $m=p$, i.~e.\ the term containing the matter structure constant~\eqref{Ok}. In the case $k=2$ we have
\begin{eqnarray}
 C_{\Delta_{1, 3},\Delta_{1,m}}^{\Delta_{1,m}}=\frac{\Gamma(2-2\rho)}{\Gamma(2\rho)}
\bigg(\frac{\gamma^3(\rho)}{\gamma(3\rho-1)}\bigg)^{\frac 12}
\frac{\gamma(1+(1-m)\rho)}{\gamma(2-(1+m)\rho)}
\end{eqnarray}
In the Liouville sector we apply the shift relations for the Upsilon function to find 
\begin{eqnarray}
C_{3b/2,Q/2+i P}^{Q/2+i P}=(\pi\gamma(b^2)b^{2-2b^2})^{-3/2}
4 \frac{\Upsilon(b) \Upsilon(3 b)}{\Upsilon^2(3b/2)} \frac{P^2 \Upsilon(b+2 i P)\Upsilon(b-2 i P)}
{\Upsilon(3b/2+2 i P)\Upsilon(3b/2-2 i P)}
\end{eqnarray}
and
\begin{eqnarray}
C_{2b,Q/2+i P}^{Q/2+i P}=(\pi\gamma(b^2)b^{2-2b^2})^{-2}4 b^{-1-2b^2} \gamma(3b^2)
\frac{\Upsilon(b) \Upsilon(3 b)}{\Upsilon^2(2b)} \frac{P^2}
{\gamma(b^2+2 i b P)\gamma(b^2-2i b P)}
\end{eqnarray} 
where the explicit integral representations for the combinations of the Upsilon functions are
\begin{eqnarray}
&\frac{\Upsilon(b) \Upsilon(3 b)}{\Upsilon^2(3b/2)}=\exp\bigg\{-\int_0^{\infty}\frac{d t}{2t}
\bigg[
\frac{
\cosh\big(\frac{t-b^2 t}{2b}\big)-2\cosh\big(\frac{t-2b^2 t}{2b}\big)
+\cosh\big(\frac{t-5b^2 t}{2b}\big)}
{\sinh\big(\frac{b t}{2}\big)\sinh\big(\frac{t}{2b}\big)}
+(2-9b^2)e^{-t}
\bigg]\bigg\}
\nonumber\\
&\frac{\Upsilon(b+2 i P) \Upsilon(b-2 i P)}{\Upsilon(3b/2+2 i P)\Upsilon(3b/2-2 i P)}
=\exp\bigg\{-\int_0^{\infty}\frac {d t}{2 t}
\bigg[2\frac{\cosh\big(2 P t \big) \sinh\big(\frac{2t-3b^2t}{2b}\big)}
{\sinh\big(\frac{t}{2b}\big)\cosh\big(\frac{bt}{4}\big)}
-(2-3b^2)e^{-t}\bigg]\bigg\}\\
&\frac{\Upsilon(b) \Upsilon(3 b)}{\Upsilon^2(2b)}=\exp\bigg\{-\int_0^{\infty}\frac {d t}{t}
\bigg[2\cosh\big(\frac{t-3b^2 t}{2b}\big)
\sinh\big(\frac{b t}{2}\big)\sinh^{-1}\big(\frac{t}{2b}\big)-2b^2e^{-t}\bigg]\bigg\}\nonumber
\end{eqnarray}
These representations of the Liouville structure constants are applicable for
the values of the parameter $b$ related as in~\eqref{bp} to arbitrary 
positive $p$.

\section{Comparing with Matrix Models}
The Liouville gravity and the matrix models approaches are expected to be physically equivalent 
since they arise from the same idea of fluctuating $2D$ geometries. There exist numerous 
confirmations of this 
idea (see e.g.~\cite{KPZ,GoulLi,DiFrK}). In~\cite{Moore} the equivalence of the 
minimal gravity $\mathcal{MG}_{2/2p+1}$ with the $p$-critical one-matrix models
was verified up to the level of two-point correlation functions. 
This comparison is not 
straightforward due to the so called resonance 
ambiguity. In~\cite{BZ} the special resonance  
transformation, which relates the coupling parameters of the Liouville gravity
to the parameters describing the deviation from the $p$-critical point in the matrix models,
was proposed. This allows one, in principle, to identify $n$-correlation functions. This conjecture was 
checked up to four-point correlation numbers on the sphere. In~\cite{BT} the resonance transformation
was applied to find the generating function of the correlation numbers on the torus and
to compute some correlators in the coordinates corresponding to the minimal Liouville gravity.
Here we brifly summarize these results. The genus one  
contribution in the partition function of the one-matrix model is
\begin{equation}
Z_{\text{torus}}=-\frac{\ln P'(u_*)}{12}
\end{equation} 
where in the Liouville frame the string polynomial is defined as
\begin{equation}
P(u)=\frac{L_{p+1}(u)-L_{p-1}(u)}{2p+1} + 
\sum_{n=1}^{\infty}\sum_{k_1,\cdots,k_n=0}^{p-1} \frac{\lambda_{k_1}\cdots\lambda_{k_n}}{n!} 
\frac{d^{n-1}}{d x^{n-1}} L_{p-\sum k_i-n}(u) 
\end{equation}
Here $L_k(u)$ are the Legendre polynomials and $u_*$ is the maximal real root of $P(u)$. The correlation 
numbers are expressed as
\begin{equation}
\langle O_{k_1} \cdots O_{k_n} \rangle_{\text{torus}} = \frac{\partial^n Z_{\text{torus}}}{\partial\lambda_{k1}\cdots\partial\lambda_{k_n}}\bigg{|}_{\lambda_1=\cdots=\lambda_{p-1}=0}
\end{equation}
In particular, for the one-point amplitude one can derive
\begin{eqnarray}
\langle O_k\rangle_{\text{torus}}=\frac{d}{d \lambda_k} Z_{\text{torus}}(u_*)
=\frac{(2 p-k)(k+1)}{24}\label{OkMM}
\end{eqnarray}

Now we are going to test our expression~\eqref{Ok} against the matrix models correlation numbers~\eqref{OkMM}. 
Since the normalization of the partition function cannot be fixed in a universal way it is natural
to relate the ratio of two different correlation numbers. Moreover one should take into account the 
different normalization of the operators in the Liouville gravity and the matrix model approach. 
The normalization of the operators does not depend 
on the topology and can be adopted from the caclulation on the sphere~\cite{BAlZ}:
\begin{equation}
O_{k}^{\text{MM}}= N(a_{1,-k-1}) O_k^{\text{MLG}}
\end{equation}
where
\begin{equation}
N(a)=(\pi \gamma(b^2))^{a/b} \bigg(\gamma(2 a b- b^2)\gamma(2a/b-1/b^2)\bigg)^{-1/2}
\end{equation}
We conclude that the following relation between torus one-point correlation numbers in the matrix models 
and in the minimal Liouville gravity takes place
\begin{equation}
\frac{N(a_{1,-k-1})}{N(b)}\frac{\langle O_k \rangle^{\text{MLG}}_{\rm torus}}
{\langle O_0 \rangle^{\text{MLG}}_{\rm torus}}
=\frac{\langle O_k \rangle^{\text{MM}}_{\rm torus}}{\langle O_0 \rangle^{\text{MM}}_{\rm torus}}
\label{nrmlzO13} 
\end{equation}
We analyzed numerically two examples $k=(1,2)$. For the correlation function 
$\langle O_0 \rangle^{\text{MLG}}$ we used the exact result~\eqref{1pointexact}, while for the 
correlation number in the numerator we used the approximative expression~\eqref{Ok}. 
From~\eqref{OkMM} for $k=1$ it follows
\begin{equation}
\frac{N(3b/2)}{N(b)}\frac{\langle O_1 \rangle^{\text{MLG}}_{\rm torus}}
{\langle O_0 \rangle^{\text{MLG}}_{\rm torus}}
=\frac{2p-1}{p}
\label{result1} 
\end{equation}
and for $k=2$
\begin{equation}
\frac{N(2b)}{N(b)}\frac{\langle O_2 \rangle^{\text{MLG}}_{\rm torus}}
{\langle O_0 \rangle^{\text{MLG}}_{\rm torus}}
=\frac{3(p-1)}{p}
\label{result2} 
\end{equation}
We checked relations~\eqref{result1} and~\eqref{result2} for different models 
$\mathcal{MG}_{2/2p+1}$ with $p=(1,\cdots,20)$ and we have found that the results 
match with very good accuracy
(always about five digits). For example, for $p=7$ we find the left hand side of~\eqref{result1}
is equal $1.85715$ while the right hand side is $1.85714$ and the left hand side of~\eqref{result2}
is equal $2.57134$ while the right hand side is $2.57143$.

\section{Discussion}
We have verified that the resonance transformation proposed in~\cite{BZ} allows one to
relate the matrix models correlation functions with those of the minimal Liouville gravity for higher
genera topologies. The conformal block expansion that enters the expression for the one-point correlation 
functions converges very fast. In fact, it is sufficient to retain the zero-order contribution in the conformal 
blocks to have a great numerical accuracy. 
Nevertheless, a derivation of the analytic answer for the torus amplitudes by means of minimal Liouville gravity 
methods is still missing.  
We suppose that the higher equations of motion in the Liouville theory~\cite{HEM} are relevant 
to this task as it was the case for the spherical topology. It was shown in~\cite{BBSusy}
that the following consequence of higher equations of motion takes place 
\begin{eqnarray}
B_{m,n} O_{m,n}= Q \bar Q \mathbb{O}'_{m,n} 
\end{eqnarray}
where $Q$ is the BRST charge, $B_{m,n}$ are some numeric factors and $\mathbb O'_{m,n}$ are the logarithmic counterparts 
of the so called ground ring physical 
fields $\mathbb O_{m,n}$ 
(see~\cite{BBSusy,BAlZ} for more details). Taking into account the commutation relation 
$[b_k,Q]_+=L_k$ one can conclude that the integrand in~\eqref{1pointB} should have the form of a total derivative with 
respect to the period $\tau$
\begin{eqnarray}
\langle O_{m,n} \rangle_{\rm torus}= B_{m,n}^{-1}\int_{F} d\tau\, d\bar{\tau}\,
\partial_{\tau}\partial_{\bar\tau} \langle  \mathbb{O}'_{m,n} \rangle_{\rm torus} 
\end{eqnarray}
The one-point amplitude is hence defined by the asymptotic behavior of the correlation function 
$\langle\mathbb{O}'_{m,n} \rangle_{\rm torus}$ near the boundary of the moduli space.  
At present this remains a conjecture.

\section*{Acknowledgments}
The author thanks A.~Belavin, S.~Rebault, V.~Fateev and M.~Lashkevich for useful discussions. Also he 
acknowledges the hospitality of the LPTA of University II of Montpellier. The research was
supported by Federal Program Scientific-Pedagogical Personnel of Innovation Russia  
(State Contract No.~02.740.11.5165) and by RFBR (the initiative interdisciplinary project 
Grant No.~09-02-12446-ofi-m and Grant No.~08-01-00720-a)
and also by an RFBR-CNRS project (Grant PICS No.~09-02-91064).

\end{document}